\documentclass[twocolumn,aps,prb,showpacs]{revtex4}

\usepackage{epsfig}

\begin{document}


\title{Impurity resonance states in noncentrosymmetric superconductor $CePt_{3}Si$: a probe for Cooper-pairing symmetry}
\author{Bin Liu$^1$, and Ilya Eremin$^{1,2}$ }

\affiliation{$^1$ Max-Planck-Institut f\"ur Physik komplexer
Systeme, D-01187 Dresden, Germany \\
$^{2}$ Institute f\"{u}r Mathematische und Theoretische Physik,
TU-Braunschweig, D-38106 Braunschweig, Germany}

\begin{abstract}
Motivated by the recent discovery of noncentrosymmetric
superconductors, such as $CePt_{3}Si$, $CeRhSi_{3}$ and
$CeIrSi_{3}$, we investigate theoretically the impurity resonance
states with coexisting  $s$- and $p$-wave pairing symmetries. Due to
the nodal structure of the gap function, we find single nonmagnetic
impurity-induced resonances appearing in the local density of state
(LDOS). In particular, we analyze the evolution of the local density
of states for coexisting isotropic $s$-wave and $p$-wave
superconducting states and compare with that of anisotropic $s$-wave
and $p$-wave symmetries of the superconducting gap. Our results show
that the scanning tunneling microscopy can shed light on the
particular structure of the superconducting gap in
non-centrosymmetric superconductors.
\end{abstract}

\maketitle

\section{ Introduction}

Recent discoveries of superconductivity in the systems that posses a
lack of inversion symmetry such as $CePt_{3}Si$\cite{bauer} with
$T_{c}\simeq0.75K$ and more recently $CeRhSi_{3}$\cite{kimura},
$CeIrSi_{3}$\cite{sugitani},
$Li(Pd_{1-x},Pt_{x})_{3}B$\cite{togano}, $UIr$\cite{akazawa},
$Y_{2}C_{3}$\cite{amano} have raised an interest in the theoretical
investigation of superconductivity in these systems. Among
interesting questions the most important one concerns the underlying
symmetry of the superconducting order parameter. In particular, in
all these materials, there is a nonzero potential gradient $\nabla
V$ averaged in the unit cell due to lack of inversion symmetry,
which results in the anisotropic spin-orbit interaction. Its general
form can be determined by a group theoretical
argument\cite{samokhin} and, as it has been found, leads to many
interesting properties
\cite{edel,rashba,yip,sigrist,fujimoto1,yanase,eremin}. For example,
on general grounds there is a mixing of the spin-singlet and spin
triplet superconducting states due to the lack of inversion. In
CePt$_3$Si the pairing symmetry has been studied
theoretically\cite{rashba,yip,sigrist,fujimoto1,yanase} and it is
believed that the $s + p$-wave superconducting state is realized.
Frigeri {\it et al.}\cite{sigrist} pointed out that the spin-orbit
interaction could determine the direction of the ${\bf d}$-vector as
${\bf d} || \vec{l}$ ($\vec{l}$ is the vector of the Rashba
spin-orbit coupling) for which the highest transition temperature
was obtained. A microscopic calculation with the detailed structure
of the Fermi surface\cite{yanase} seems to confirm that the $s+p$
wave state is the most probable one. The experimental studies of the
temperature dependencies of the spin-lattice relaxation\cite{yogi},
the magnetic penetration depth\cite{bonalde}, and the thermal
conductivity measurements\cite{izawa} are also consistent with this
conjecture.

It is known that the non-magnetic as well as the magnetic impurities
in the conventional and unconventional superconductors already have
been proven to be a useful tool to distinguish between various
symmetries of the superconducting state\cite{balatskyREV}. For
example, in the conventional isotropic $s$-wave superconductor the
single magnetic impurity induced resonance state is located at the
gap edge, which is known as Yu-Shiba-Rusinov state\cite{yu}. In the
case of unconventional superconductor with $d_{x^{2}-y^{2}}$-wave
symmetry of the superconducting state the non-magnetic
impurity-induced bound state appears near the Fermi energy as a
hallmark of $d_{x^{2}-y^{2}}$-wave pairing symmetry\cite{balatsky}.
The origin of this difference is understood as being due to the
nodal structure of two kinds of SC order: in the
$d_{x^{2}-y^{2}}$-wave case the phase of Cooper-pairing wave
function changes sign across the nodal line which yields finite
density of states below the superconducting gap, while in the
isotropic $s$-wave case the density of states is gapped up to
energies of about $\Delta_0$ and thus the bound state can appear
only at the gap edge. In principle the formation of the impurity
resonance states can also occur in unconventional superconductors if
the nodal line or point does not exist at the Fermi surface of a
superconductor like it occurs for isotropic nodeless $p$-wave and/or
$d_{x}+id_{y}$-wave superconductors for the large value of the
potential strength\cite{wang}. Therefore, STM measurements of the
impurity states can provide important messages about the pairing
symmetry in the revelent systems. In the noncentrosymmetric
superconductor with the possible coexistence of s-wave and p-wave
pairing symmetry, it is very interesting to see what is the nature
of the impurity state, and whether a low energy resonance state can
still occur around the impurity through changing the dominant role
played by each of the pairing components. Previously the effect of
the non-magnetic impurity scattering has been studied in the
non-centrosymmetric superconductors with respect to the suppression
of T$_c $\cite{frigeriEPB} and the behavior of the upper critical
field\cite{minsam}.

In this paper we investigate theoretically the impurity resonance
states where both $s$-wave and $p$-wave Cooper-pairing coexist. Due
to the nodal structure of gap function as a result of the
interference between the spin triplet and the spin singlet
components of the superconducting order parameters, we find that a
single nonmagnetic impurity-induced resonance state appears in the
local density of state.  In particular, we analyze the evolution of
the local density of states for coexisting isotropic $s$-wave and
$p$-wave superconducting states and compare with that of anisotropic
$s$-wave and $p$-wave symmetries of the superconducting gap. Our
results show that the scanning tunneling microscopy can shed light
on the particular structure of the superconducting gap in
non-centrosymmetric superconductors.

\section{the Model and T-matrix formulation}

Theoretical models of the superconducting state in CePt$_3$Si are
based upon the existence of a Rashba type spin-orbit coupling
(RSOC)\cite{rashba}. Therefore, following previous
consideration\cite{sigrist} we start from an single orbital model
with RSOC
\begin{eqnarray}
H&=&\sum_{{\bf k}s}\varepsilon_{\bf k}c_{{\bf k}s}^{\dagger}c_{{\bf
k}s}+\alpha\sum_{{\bf k}ss'}{\bf g_{k}}\cdot{\bf
\sigma_{ss'}}c_{{\bf k}s}^{\dagger}c_{{\bf k}s'}, \label{eq:1}
\end{eqnarray}
where $c_{{\bf k}s}^{\dagger}$ ($c_{{\bf k}s}$) is the fermion
creation (annihilation) operator with spin $s$ and momentum ${\bf
k}$. Here, $\varepsilon_{\bf k}$ is the tight-binding energy
dispersion
\begin{eqnarray}
\varepsilon_{\bf
k}&=&2t(\cos(k_{x})+\cos(k_{y}))+4t_{1}\cos(k_{x})\cos(k_{y})
 \nonumber\\
&+&2t_{2}(\cos(2k_{x})+\cos(2k_{y}))\nonumber\\
&+&[2t_{3}+4t_{4}(\cos(k_{x})+\cos(k_{y}))\nonumber\\&+&4t_{5}(\cos(2k_{x})+\cos(2k_{y}))]\cos(k_{z})\nonumber\\
&+&2t_{6}\cos(2k_{z})-\mu
\end{eqnarray}
which reproduces the so-called $\beta$-band of $CePt_{3}Si$ as
obtained from the band structure calculations\cite{samokhin,yanase}.
The electron hopping parameters are
$(t,t_{1},t_{2},t_{3},t_{4},t_{5},t_{6},n)=(1,-0.15,-0.5,-0.3,-0.1,-0.09,-0.2,1.75)$
and the electron density per site $n$ is used to determine the
chemical potential\cite{yanase}.

\begin{figure}[ht]
\epsfxsize=3.5in\centerline{\epsffile{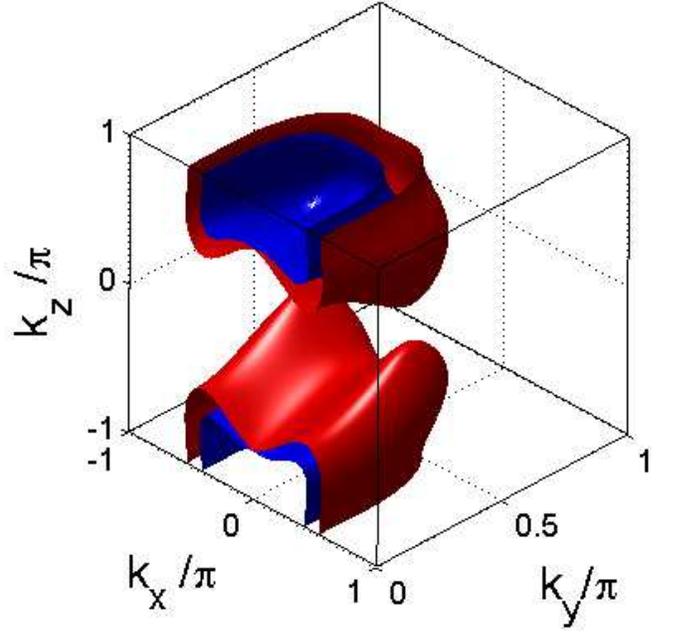}} \caption{ (color
online) The calculated Fermi surface using the Eq. (\ref{eq:1}) and
the spin-orbit coupling constant $\alpha=0.3t$.} \label{fig1}
\end{figure}

The second term of Eq.(\ref{eq:1}) is the RSOC interaction where
$\alpha$ denotes the coupling constant and the vector function ${\bf
g_{k}}$ is assumed in the following form ${\bf g_{k}}=(- \sin k_{y},
\sin k_{x}, 0)$. This term removes the usual Kramers degeneracy
between the two spin states at a given ${\bf k}$, and leads to a
quasiparticle dispersion $\epsilon_{\bf k}=\varepsilon_{\bf
k}\pm\alpha|{\bf g_{k}}|$ with ${|{\bf g_{k}}|=\sqrt{{\bf
g_{k}}^{2}_{x}+{\bf g_{k}}^{2}_{y}+{\bf g_{k}}^{2}_{z}}}$, splitting
the Fermi surface (FS) into two sheets. Based on the above hopping
parameters and RSOC constant $\alpha=0.3t$, the resulting FS is shown
in Fig.\ref{fig1}, where the main characteristic features of the FS
has been successfully reproduced\cite{samokhin}.

In the superconducting state, the presence of RSOC breaks the parity
and, therefore, mixes the singlet (even parity) and triplet (odd
parity) Cooper-pairing states. A full symmetry
analysis\cite{yanase,samokhin} shows that $s$-wave pairing
$\Delta_{s}=\Delta_{0}(\cos(k_{x})+\cos(k_{y}))$ and a $p$-wave
triplet pairing state with order parameter ${\bf d_{k}}$ parallel to
the ${\bf g_{k}}$ vector, ${\bf d_{k}}=d_{0}{\bf g_{k}}$ are able to
coexist. Following previous estimations\cite{yanase} we have taken
the odd parity component ${\bf d_{k}} =d_{0}(- \sin k_y, \sin k_x,
0)$. Then the mean field BCS Hamiltonian for this system has the
matrix form
\begin{eqnarray}
H_{\bf k}=\left (\matrix{\varepsilon_{\bf k} &\alpha g &-d^{\ast}
&\Delta_{\bf k}\cr \alpha g^{\ast} &\varepsilon_{\bf k}
&-\Delta_{\bf k} &d\cr -d &-\Delta^{\ast}_{\bf k} &-\varepsilon_{\bf
k} &\alpha g^{\ast}\cr \Delta^{\ast}_{\bf k} &d^{\ast} &\alpha g
&-\varepsilon_{\bf k}\cr}\right). \label{eq:BCS}
\end{eqnarray}
Where for briefly, $g=({\bf g_{k}}_{x}-i{\bf g_{k}}_{y})$ and
$d=({\bf d_{k}}_{x}+i{\bf d_{k}}_{y})$.
The inverse of the single-particle Green's function is defined as
\begin{eqnarray}
g^{-1}({\bf k},i\omega_{n})=i\omega_{n}I-H_{\bf k},
\end{eqnarray}
where $I$ is the $4\times4$ identity matrix. Taking the inverse of
Eq. (\ref{eq:BCS}) we find
\begin{eqnarray}
g({\bf k},i\omega_{n})=\left (\matrix{G({\bf k},i\omega_{n}) &F({\bf
k},i\omega_{n})\cr F^{\dagger}({\bf k},i\omega_{n}) &-G^{t}(-{\bf
k},-i\omega_{n})\cr}\right)
\end{eqnarray}
where
\begin{mathletters}
\begin{eqnarray}
G({\bf k},i\omega_{n})&=&\sum_{\tau=\pm1}\frac{1+\tau({\bf
\vec{g}_{{\bf k}}}\cdot{\bf
\sigma})}{2}G_{\tau}({\bf k},i\omega_{n}), \\
F({\bf k},i\omega_{n})&=&\sum_{\tau=\pm1}\frac{1+\tau({\bf
\vec{g}_{\bf k}}\cdot{\bf \sigma})}{2}i\sigma_{y}F_{\tau}({\bf
k},i\omega_{n}),
\end{eqnarray}
\end{mathletters}
and
\begin{mathletters}
\begin{eqnarray}
G_{\tau}({\bf k},i\omega_{n})&=&\frac{i\omega_{n}+\epsilon_{\tau}}{(i\omega_{n})^{2}-E^{2}_{{\bf k}\tau}}, \\
F_{\tau}({\bf
k},i\omega_{n})&=&\frac{\Delta_{\tau}}{(i\omega_{n})^{2}-E^{2}_{{\bf
k}\tau}}.
\end{eqnarray}
\end{mathletters}
Here, the single-particle excitation energy is
\begin{eqnarray}
E_{{\bf k}\tau}=\sqrt{\epsilon_{\tau}^{2}+|\Delta_{\tau}|^{2}}
\end{eqnarray}
with
\begin{eqnarray}
\epsilon_{\tau}=\varepsilon_{\bf {\bf k}}+\tau\alpha|{\bf g_{k}}|;
\Delta_{\tau}=\Delta_{\bf k}+\tau|{\bf d_{k}}|,
\end{eqnarray}
and the unit vector is ${\bf \vec{g}_{\bf k}}={\bf g_{\bf k}}/{|{\bf
g_{\bf k}}|}$.

For completion the equations above have to be supplemented by the
self-consistency equation that determines the symmetry of the
superconducting gap and the superconducting transition temperature.
For the sake of simplicity and also because this is not critical for
our further analysis we consider the superconducting order parameter
as a given parameter. At the same time, recent studies based on the
helical spin fluctuation mediated Cooper-pairing find two stable
superconducting phases with either dominantly $s+p$-wave or
$p+d+f$-wave symmetry of superconducting order
parameter\cite{yanase,sigrist08,fujimoto}. In the following we adopt
the former one for our calculation.

The next step is to obtain the Green's function in the presence of
a single impurity site. The impurity scattering is given by
\begin{eqnarray}
H_{imp}=U_{0}\sum_{\sigma}c_{0\sigma}^{\dagger}c_{0\sigma},
\end{eqnarray}
where without loss of generality we have taken a single-site
nonmagnetic impurity of strength $U_{0}$ located at the origin,
$r_{i}=0$. Then the site dependent Green's function can be written
in terms of the T-matrix formulation\cite{wang,fisher,wang1} as
\begin{eqnarray}
\zeta(i,j;i\omega_{n})&=&\zeta_{0}(i-j;i\omega_{n}) \nonumber\\
&+&\zeta_{0}(i,i\omega_{n})T(i\omega_{n})\zeta_{0}(j,i\omega_{n}),
\end{eqnarray}
where
\begin{eqnarray}
T(i\omega_{n})&=&\frac{U_{0}\rho_{3}}{1-U_{0}\rho_{3}\zeta_{0}(0,0;i\omega_{n})}\\
\zeta_{0}(i,j;i\omega_{n})&=&\frac{1}{N}\sum_{\bf k}e^{i\bf
k\cdot\bf R_{ij}}g(k,i\omega_{n}),
\end{eqnarray}
with $\rho_{i}$ being the Pauli spin operator, and $\bf R_{i}$ is
the lattice vector, $\bf R_{ij}=\bf R_{i}-\bf R_{j}$. Finally, the
local density of state  which can be measured in the STM experiment
has been obtained as
\begin{eqnarray}
N(r,\omega)=-\frac{1}{\pi}\sum_{i}{\rm
Im}\zeta_{ii}(r,r;\omega+i\eta),
\end{eqnarray}
where $\eta$ denotes an infinitely small positive number.

\section{Numerical Results and Discussions}

\subsection{The density of state}

Before considering the effect of the impurity it is useful to
analyze first the density of state (DOS) in the superconducting
state, which is expressed as,
\begin{eqnarray}
\rho(\omega)=-\frac{1}{\pi}{\rm Im}\sum_{i, {\bf k}}g_{ii}({\bf
k},\omega)
\end{eqnarray}
\begin{figure}[ht]
\epsfxsize=3.5in\centerline{\epsffile{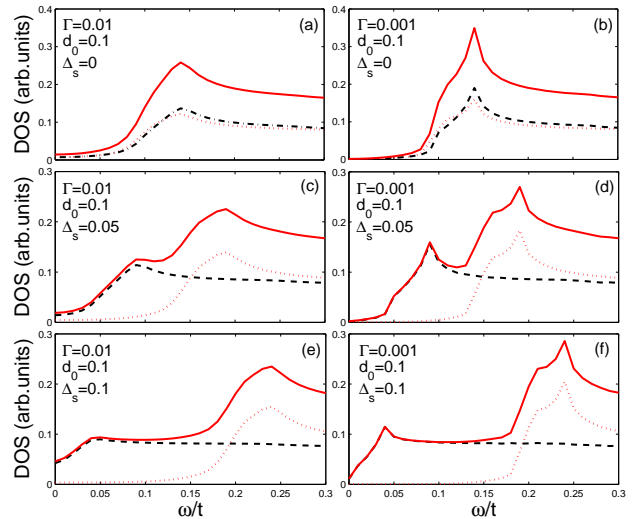}} \caption{ (color
online) The evolution of the local density of states for various
ratio between coexisting isotropic $s$-wave and $p$-wave
Cooper-pairing state. The left and right panels refer to the
different values of the damping constant $\Gamma$. The dashed and
the dotted curve denote the contribution of the different bands and
the straight curve refers to the total density of states. The
parameters of the gaps and the damping $\Gamma$ are given in terms
of hopping integral $t$.} \label{fig2}
\end{figure}
As we already have mentioned above it is not necessary to calculate
the magnitude of the gap functions self-consistently since we are
mainly interested in the qualitative properties arising from the gap
structure. We first consider the situation when the $s$-wave part of
the total superconducting gap is momentum independent, $\Delta_{s} =
\Delta_0$. In Fig.\ref{fig2} we show the evolution of the density of
state for positive frequencies for various values of the $s$-wave
component of the superconducting gap. In particular, for zero value
of the $s$-wave component the superconducting gap is purely
determined by the $p$-wave superconducting gap with the point node
at the Fermi surfaces of the corresponding bands at ($k_x=0,
k_y=0$). This gap structure is the same for both bands splitted by
the spin-orbit coupling. With increasing value of the isotropic
$s$-wave gap one finds that the total superconducting gap in one of
the bands increasing  with the total superconducting gap $\Delta_{s}
+|{\bf d}_{\bf k}|$ while it decreases effectively for the other
band for which the total gap is $\Delta_{s} - |{\bf d}_{\bf k}|$.
Once both $s-$wave and $p$-wave superconducting gaps are the same,
the accidental node forms at one of the band and the behavior of the
density of states changes to a linear at low energy reflecting  the
formation of the line of node. We further note that density of
states shows only slight electron-hole asymmetry.

\begin{figure}[ht]
\epsfxsize=3.5in\centerline{\epsffile{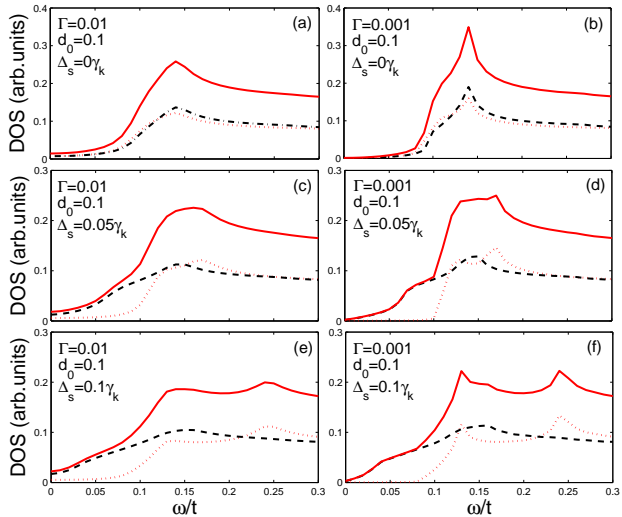}} \caption{ (color
online) The evolution of the local density of states for various
ratio between coexisting anisotropic $s$-wave and $p$-wave
Cooper-pairing state. The left and right panels refer to the
different values of the damping constant $\Gamma$. The dashed and
the dotted curve denote the contribution of the different bands and
the straight curve refers to the total density of states. The
parameters of the gaps and the damping $\Gamma$ are given in terms
of hopping integral $t$.} \label{fig3}
\end{figure}

In Fig. \ref{fig3} we show a similar evolution of the density of
states, however, now the $s$-wave component of the superconducting
gap is momentum dependent, $\Delta_{s} = \Delta_0 (\cos k_x +\cos
k_y) = \Delta_0 \gamma_{\bf k}$ \cite{comment}.
Interestingly enough, here the node in the density of states forms
already when the $p$-wave superconducting gap component is zero (see
also Fig.\ref{fig5}) and is the result of the initial momentum
structure of the $s$-wave superconducting gap that yields point
nodes on the Fermi surface. This is unique to the anisotropic
$s$-wave superconducting gap. By introducing the interference
between $s$-wave and $p$-wave gap the position of the node is
shifted to the different points of the Brillouin Zone, however, here
the nodal structure of the superconducting gap is not a result of
the interference effect between $p$-wave and $s$-wave of the
superconducting gap but arises already in the pure anisotropic
$s$-wave symmetry and shifted by introducing the moderate component
of the $p$-wave gap.

\subsection{Impurity resonance states}
\begin{figure}[ht]
\epsfxsize=3.5in\centerline{\epsffile{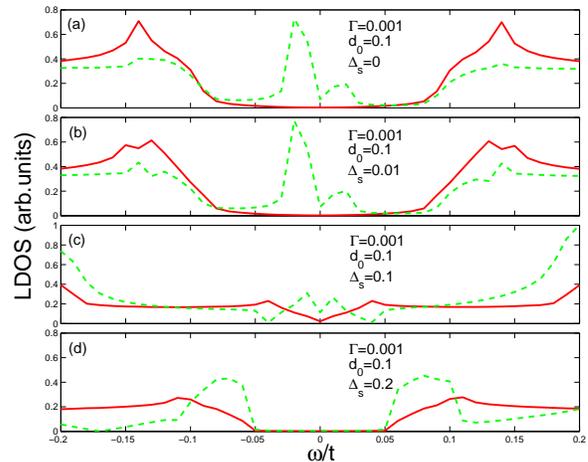}} \caption{
(color online) The LDOS for coexisting isotropic $s$-wave and
$p$-wave Cooper pairing states for various ratio of the parameters.
The straight (red) curves refer to the calculated density of states
without impurity and the dashed (green) curves refer to the LDOS at
the $(0,1,0)$ position. Here, we use $U_0=5t$.} \label{fig4}
\end{figure}
\begin{figure}[ht]
\epsfxsize=3.5in\centerline{\epsffile{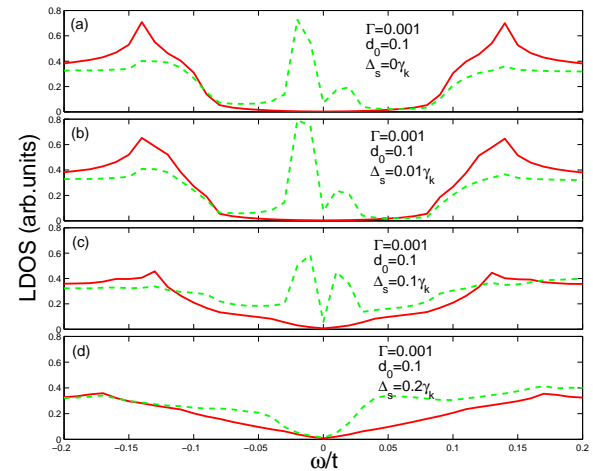}} \caption{
(color online) The LDOS for coexisting anisotropic $s$-wave
($\Delta_{\bf }$)and $p$-wave Cooper pairing states for various
ratio of the parameters. The straight (red) curves refer to the
calculated density of states without impurity and the dashed (green)
curves refer to the LDOS at the $(0,1,0)$ position. Here, we use
$U_0=5t$.} \label{fig5}
\end{figure}
In view of complicated band structure arising in CePt$_3$Si from the
Rashba spin-orbit coupling and the corresponding interference effect
for the superconducting gap the density of states in a clean case
that can be accessed by the tunneling experiments cannot give a
precise information on the exact structure of the superconducting
gap in the non-centrosymmetric superconductors. At the same time, an
introduction of the non-magnetic impurity can give additional
important information on the symmetry of the superconducting gap in
such a material. In terms of Eq.(16), the T-matrix can be written as
\begin{eqnarray}
T^{-1}(i\omega_{n})=U^{-1}_{0}-\rho_{3}\zeta_{0}(0,0;i\omega_{n}),
\end{eqnarray}
and the position of the impurity resonant state is given by $\det
T^{-1}=0$.
We first study the situation of the isotropic $s$-wave
superconducting gap coexisting with $p$-wave. In Fig.\ref{fig4} we
show the calculated density of states without impurity and also the
local density of states with an impurity on the nearest neighbor
site $(0,1,0)$. Without the $s$-wave component the density of states
shows the formation of the impurity induced resonant bound states
that appear symmetrically in energy at the positive and negative
sides of the LDOS. Clearly these resonant bound states arise due to
unconventional nature of the $p$-wave superconducting gap and the
nodal points at the Fermi surface. One clearly sees that upon
increasing of the isotopic $s$-wave contribution the bound state
shifts towards the edge of the superconducting gap implying the zero
density of states for energies lower than $\Delta_0$.

In Fig.\ref{fig5} we show the corresponding local density of states
for the coexisting anisotropic $s$-wave and $p$-wave superconducting
gaps.In the present case, for any value of the $s$-wave and $p$-wave gap
there are nodal points at the Fermi surface resulting either from
the internal structure of the anisotropic $s$-wave gap, point nodes
from the $p$-wave state, or a nodal line at one of the bands that
arises due to interference of the $p$-wave and $s$-wave gap.
Therefore, the impurity induced bound state occurs for all ratios
between the $p$-wave and $s$-wave gap. Note, that in case of pure
anisotropic $s$-wave gap due to the nodal structure on both of the
bands the impurity induced bound state becomes visible only for a
very large values of the potential scattering strength $U_0$.

\section{Summary}

In summary, we have investigated theoretically the non-magnetic
impurity induced resonance bound states in the superconductors
without inversion symmetry using as an example $CePt_{3}Si$, which
is believed to have a line node in the energy gap arising from the
coexistence of $s$-wave and $p$-wave pairing symmetry. Analyzing the
local density of states we find that the nodal structure of gap
function, we find that a single nonmagnetic impurity-induced
resonance states is highly probable in non-centrosymmetric
superconductors. We show that further STM experiments may reveal the
exact symmetry of the superconducting gap in these systems.

We thank Jun Chang for useful discussions.


\begin{references}
\bibitem{comment} Note that both the nodal and the isotropic $s$-wave components of
the total superconducting gap may coexist in the microscopic theory
of the Cooper-pairing. Moreover, the on-site Coulomb repulsion  will
tend to favor the nodal $s$-wave gap function.
\end{references}
\end{document}